\documentclass[aps,prb,showpacs,amsmath,amssymb,amsfonts,twocolumn,superscriptaddress,footinbib]{revtex4}

\usepackage{afterpage}
\usepackage{float}

\usepackage{subfigure}
\usepackage{graphicx}
\usepackage[flushmargin]{footmisc}
\usepackage{color}

\begin{document}

\title{Suppression of Zeeman splitting and polarization steps \\
       in localized exciton-polariton condensates}

 \author{T. C. H. Liew}
 \affiliation{School of Physics and Astronomy, University of Southampton, Highfield, Southampton SO17 1BJ, UK}
 \author{Yuri G. Rubo}
 \affiliation{School of Physics and Astronomy, University of Southampton, Highfield, Southampton SO17 1BJ, UK}
 \affiliation{Centro de Investigaci\'on en Energ\'{\i}a, Universidad Nacional Aut\'onoma de M\'exico, Temixco, Morelos, 62580, Mexico}
 \author{I. A. Shelykh}
 \affiliation{International Center for Condensed Matter Physics, Universidade de Brasilia, 70904-970, Brasilia-DF, Brazil}
 \affiliation{St. Petersburg State Polytechnical University, 195251, St. Petersburg, Russia}
 \author{A. V. Kavokin}
 \affiliation{School of Physics and Astronomy, University of Southampton, Highfield, Southampton SO17 1BJ, UK}
 \affiliation{Marie-Curie Chair of Excellence ``Polariton devices'', University of Rome II, 1, via della Ricerca Scientifica, Rome, 00133, Italy}

\pacs{78.67.-n, 71.36.+c, 42.55.Sa, 42.25.Kb}

\date{\today}

\begin{abstract}

\noindent We show that the condensation of exciton-polaritons in
semiconductor microcavities in an applied magnetic field manifests
itself in the quenching of the Zeeman splitting of an elliptically
polarized condensate. The circular polarization degree of a
localized condensate with a finite number of particles increases as
a function of the magnetic field with a step-like behavior. The
width of each polarization step is fixed by the polariton-polariton
interaction constants and the number of steps is fixed by the number
of polaritons in the condensate. The magnetic susceptibility of the
condensate depends qualitatively on the parity of the number of
polaritons.
\end{abstract}

\maketitle


\section{Introduction} The Bose-Einstein condensation (BEC) of
exciton-polaritons in semiconductor microcavities has been recently
reported \cite{Kasprzak2006, Balili2007}. This discovery opens the
way to the realization of coherent matter states at unusually high
temperatures and multiple applications, including polariton lasers
\cite{Butov2007}. An essential peculiarity of polaritons in planar
semiconductor microcavities containing quantum wells consists of the
presence of a polarization degree of freedom. A linear polarization
is expected to build up spontaneously as a result of polariton
BEC\cite{Laussy2006} or in the superfluid phase \cite{Shelykh2006}.
This effect can be understood by assigning the pseudospin $+1(-1)$
to each polariton, which corresponds to the right(left) circular
polarization. Polaritons with the same pseudospin repel each other,
while there is weak attraction between the polaritons with opposite
pseudospins. As a result, the energy of a polariton system is
minimized when equal numbers of left and right circularly polarized
polaritons are present. This implies a linear polarization for a
polariton system in a condensed or superfluid state.

It was recently shown that an applied magnetic field has a
pronounced effect on the polarization state of superfluid
polaritons\cite{Rubo2006}. In this case there is a competition
between the magnetic field that tries to make the polariton system
circularly polarized and the polariton-polariton interaction that
favors a linear polarization. In weak magnetic fields this
competition is resolved in the formation of an elliptically
polarized superfluid and the complete suppression of the Zeeman
splitting. Only when the magnetic field exceeds a critical value
does the polarization become circular and the Zeeman splitting
appears.

While the superfluidity of exciton-polaritons remains still a
theoretical prediction, the formation of localized polariton
condensates has been observed, both in
random\cite{Kasprzak2006,Kasprzak2007} and artificially prepared
\cite{Balili2007} traps. Localized polariton condensates were also
created by resonant optical excitation in a planar microcavity
having a sub-micron scale potential disorder~\cite{Sanvitto} and the
possibility of creating polariton condensates in micropillar
microcavities is an area of active research~\cite{Bajoni2007}.

In this paper we analyze the magnetic field effects on such a
finite-sized polariton system, or a \emph{polariton dot}, that can
contain a small number of polaritons, $N$. We assume that polaritons
behave as ideal bosons and the temperature is low enough such that
they occupy the same orbital quantum state, i.e., they form a
condensate within the dot. We consider an effect of the external
magnetic field parallel to the structure axis on the energy spectrum
and spin polarization of exciton-polaritons. We ignore for
simplicity the field effect on orbital motion of elections and holes
which results in Landau quantization. This effect has been
considered in detail by Lozovik and co-authors \cite{Lozovik}. It is
not sensitive to spin and polarization of polaritons which are in
the scope of the present work. On the other hand, we fully take into
account the Zeeman effect. We assume that each exciton polariton can
be characterized by a $g$-factor given by the exciton $g$-factor
renormalized accounting for the Hopfield coefficients \cite{Savona}.
Experimentally, the magnetic field effect on the linear optical
response of exciton-polaritons has been studied in detail
\cite{Tignon,Gibbs}, while no experimental results on polariton BEC
under magnetic field have been reported so far, to the best of our
knowledge.

\section{The polariton dot in a magnetic field}
\begin{figure}[h]
        \centering
            \subfigure{
                \put(-50,160){\parbox{4.058cm}{\sf \bf a}}{\includegraphics[width=4.058cm]{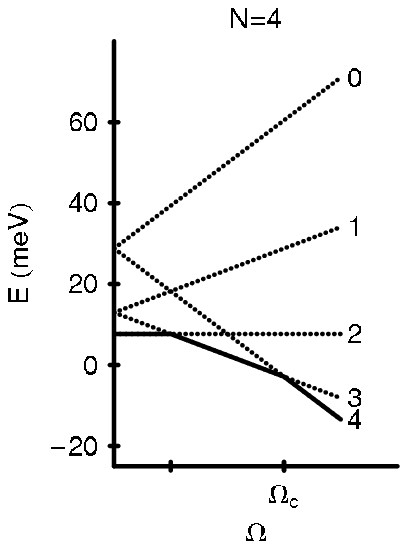}}
                }
            \subfigure{
                \put(-50,160){\parbox{4.058cm}{\sf \bf b}}{\includegraphics[width=4.058cm]{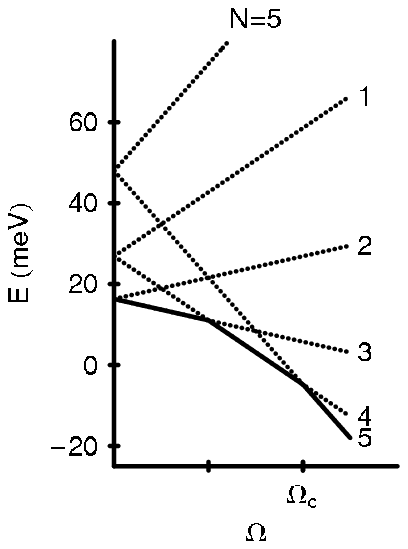}}
                }
    \caption{Dotted lines show the energy levels for an even (a)
    and odd (b) number of particles. The energy levels are labelled by the number $N_+$.
    Solid lines show the ground state. We used the parameters $\alpha_1=4.8$meV
    (assuming the dot has a diameter of about $100$nm, i.e., the dot is 20 times smaller than
    the one in Ref.~\onlinecite{Kasprzak2007})
    and $\alpha_2=-0.1\alpha_1$ (Ref.~\onlinecite{Renucci}).
    }
    \label{fig:Levels}
    \end{figure}

The Hamiltonian of our model system reads
\begin{equation}
 \label{eq:Hamilt}
 H=\frac{1}{2}\sum_{\sigma=\pm}\!\left[
 \alpha_1N_\sigma(N_\sigma-1)+\alpha_2N_\sigma N_{-\sigma}-\sigma 2\Omega N_\sigma
 \right],
\end{equation}
\noindent where the energy of non-interacting polaritons at zero
magnetic field is taken to be equal to zero, $\sigma=\pm$ is the
pseudospin index defined in such a way that $N_+$ is the number of
polaritons with spins parallel to a magnetic field and $N_-=N-N_+$
is the number of polaritons with spins antiparallel to the field.
The bare Zeeman splitting is given by $2\Omega=g \mu_B B$ where $B$
is the applied magnetic field, $g$ is the polariton g-factor, and
$\mu_B$ is the Bohr magneton. $\alpha_1$ and $\alpha_2$ are the
polariton-polariton interaction constants in triplet (parallel
spins) and singlet (anti-parallel spins) configurations,
respectively\cite{Ualpha}. The Hamiltonian (\ref{eq:Hamilt}) is
already diagonal and the energy levels $E=E(N,N_+)$ can be labeled
by two quantum numbers, $N$ and $N_+$. These levels are shown in
Fig.~\ref{fig:Levels}.

It is seen from Fig.~\ref{fig:Levels} that there are abrupt changes
of the ground state of the system as the field is increased. Each
abrupt change corresponds to the reorientation of one polariton
pseudospin, which increases the number $N_+$ by one. For an even
total number of polaritons, the ground state does not change at very
low fields and $N_+=N/2$. Then at $\Omega=(\alpha_1-\alpha_2)/2$ the
state with $N_+=(N/2)+1$ obtains a low enough energy to become the
ground state. Next, at $\Omega=3(\alpha_1-\alpha_2)/2$, the state
with $N_+=(N/2)+2$ becomes the ground state, and so on, until all
polaritons align their spin ($N_+=N$) at the critical field $B_c$
given by:
\begin{equation}
 \label{eq:Omegac}
\Omega_c=\frac{1}{2}g\mu_BB_c=\frac{1}{2}(N-1)(\alpha_1-\alpha_2).
\end{equation}

A similar behavior appears for an odd number of polaritons, however
the values of the field at which changes occur are shifted by
$(\alpha_1-\alpha_2)/2$. The critical field, when all polariton
spins are aligned, is still given by Eq.~(\ref{eq:Omegac}).

The ground state energy decreases linearly with the field, but each
change in $N_+$ causes an abrupt increase in the magnetic
susceptibility (Fig.~\ref{fig:Levels}). The overall decrease for
$\Omega<\Omega_c$ is close to parabolic for large values of $N$. For
$\Omega>\Omega_c$ the ground state energy is given by:
\begin{equation}
E_0(\Omega)=-\Omega N+\frac{\alpha_1}{2}N(N-1).
\end{equation}

\begin{figure}[h]
        \centering
            \subfigure{
                \put(-60,100){\parbox{4.058cm}{\sf \bf a}}{\includegraphics[width=4.058cm]{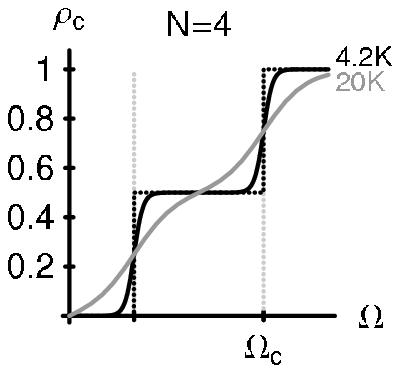}}
                }
            \subfigure{
                \put(-60,100){\parbox{4.058cm}{\sf \bf b}}{\includegraphics[width=4.058cm]{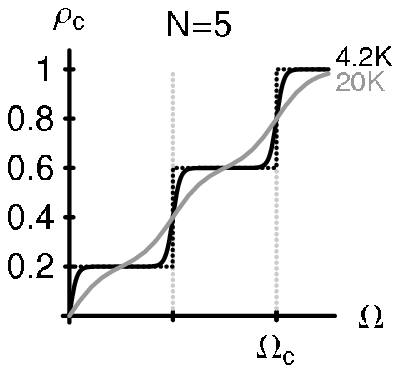}}
                }
    \caption{The dotted black line shows the circular polarization degree, $\rho_c$,
    at $T=0$ for an even (a) and odd (b) number of polaritons. The solid lines correspond to
    the finite temperatures $4.2K$ (black) and $20K$ (grey).}
    \label{fig:Steps}
    \end{figure}

These changes of the ground state lead to steps in the circular
polarization degree $\rho_c$, which is defined as:
\begin{equation}
 \label{eq:rhoc}
 \rho_c=\frac{1}{N}\left<N_+-N_-\right>=\frac{2\left<N_+\right>}{N}-1.
\end{equation}
Here the brackets indicate averaging over different states of the
polariton system. First we consider the case of a fixed number of
polaritons in the dot and assume that polaritons are in
quasi-thermal equilibrium with an effective temperature, $T$. In
this case the averaging is performed with a usual Boltzmann
distribution, i.e., the probability to observe the state with energy
$E(N,N_+)$ is:
\begin{equation}
 \label{eq:pNplus}
 p(N_+)=Z^{-1}e^{-E(N,N_+)/(k_B T)}.
\end{equation}
Here $Z$ is the partition function found from the normalization
$\sum_{N_+}p(N_+)=1$. The polarization steps are shown in
Fig.~\ref{fig:Steps}. Note that the low-field susceptibility is
sensitive to the total number of particles. At low temperatures it
is close to zero for an even number of polaritons, and it is
described by the spin-1/2 Brillouin function for an odd number of
polaritons.

There is an interesting analogy between the polarization steps we
discuss in this paper and the magnetization steps observed in the
magnetization curve of II-IV diluted magnetic semiconductors in high
magnetic fields (see, e.g., Ref.~\onlinecite{Shapira} and references
therein). The magnetization steps appear due to the change of the
ground state of antiferromagnetic pairs of Mn$^{2+}$ ions. At small
fields the ground state of a Mn$^{2+}$--Mn$^{2+}$ pair has a total
spin $S=0$ and its projection $S_z=0$. The magnetization steps
correspond to the subsequent changes of the ground state to $S=1$,
$S_z=-1$, then to $S=2$, $S_z=-2$, etc., with increasing magnetic
field, until the fully polarized state with $S=5$, $S_z=-5$ becomes
the one with the lowest energy. In general, the states of a dot with
an even number $N$ of polaritons can be (non-surjectively) mapped to
the states of an antiferromagnetic pair of two $N/4$ spins. It
should be noted, however, that the Hamiltonian's of these systems
are different. In particular, they have different numbers of
eigenstates. There are $N+1$ states in the case of a polariton dot,
and $(N+2)(N+2)/4$ states in the case of two $N/4$ spins.

As in the case of magnetization of antiferromagnetic pairs
\cite{Rubo1997}, there are several mechanisms of broadening of
polarization steps apart from the temperature broadening. Below we
consider two most important ones. Namely, these are the finite
polariton life-time $\tau_p$ and fluctuations of the number of
polaritons within the dots. Note that the finite life-time is not
negligible even if $\hbar/\tau_p$ is much less than the
polariton-polariton interaction constants. This is because each step
corresponds to the crossing of two levels with different circular
polarizations, and the broadening of polariton levels is certainly
important in the vicinity of the crossing points. The effect of
fluctuations of the polariton number is more interesting. The
positions of the steps depend only on the parity of the number of
polaritons $N$, so that both even and odd steps are expected to be
observed. On the other hand, the amplitude of each step is inversely
proportional to $N$ or $N+1$ for an even or odd number of
polaritons, respectively. Therefore, fluctuations in $N$ also
produce an additional broadening of the steps as a result of
statistical averaging.

\section{Photoluminescence in a magnetic field}
 The presence of polarization steps discussed in the previous section
is reflected in the photoluminescence (PL) spectra of the polariton
dot. In a typical PL experiment the polarization dependence of the
energy of photons emitted from the dot is observed. The energy of
the photon depends on the initial state of the polariton system and
can be written as
\begin{equation}
 \label{eq:PhotE}
 \hbar\omega=\hbar\omega_0 + \Delta E_\sigma,
\end{equation}
where $\omega_0$ is the bare single polariton frequency,
\begin{subequations}
 \label{eq:DeltaE}
 \begin{equation}
 \Delta E_+(N,N_+)=E(N,N_+)-E(N-1,N_+-1),
 \end{equation}
 \begin{equation}
 \Delta E_-(N,N_+)=E(N,N_+)-E(N-1,N_+).
 \end{equation}
\end{subequations}

\begin{figure}[h]
        \centering
            \subfigure{
                \put(-50,200){\parbox{4.058cm}{\sf \bf a}}{\includegraphics[width=4.058cm]{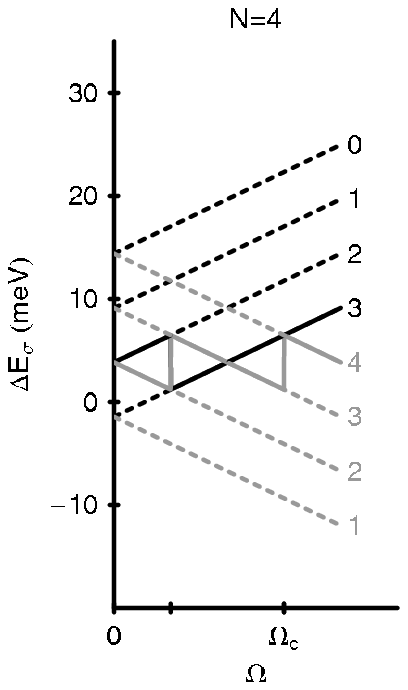}}
                }
            \subfigure{
                \put(-50,200){\parbox{4.058cm}{\sf \bf b}}{\includegraphics[width=4.058cm]{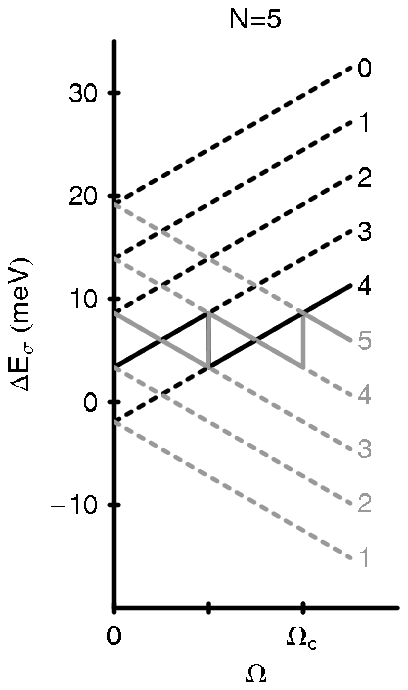}}
                }
    \caption{Black and grey lines show the energy changes, $\Delta
    E_-$ and $\Delta
    E_+$, for the emission of a photon with $\sigma_-$ and $\sigma_+$
    polarization respectively. Solid lines represent the energy changes
    from the initial ground state, whilst dashed lines show the
    possible energy changes from excited states. (a) An even
    number of polaritons in the initial state; (b) an odd number.
    Each transition is labeled by the initial value of $N_+$.}
    \label{fig:DeltaE}
    \end{figure}

The energy differences $\Delta E_\sigma$ are shown in
Fig.~\ref{fig:DeltaE}. It is seen that there is a pronounced
suppression of the Zeeman splitting of the emission from the ground
state for magnetic fields below the critical field, $\Omega_c$. A
noticeable Zeeman splitting remains, however, for the excited
states. Therefore, it is important to analyze the behavior of PL
lines at a finite effective temperature.

To calculate the emission spectrum we introduce a linewidth
broadening defined by $\Gamma=\hbar/\tau_p$. Then, the probability
of the system to emit a $\sigma$-polarized photon with energy
$\hbar\omega=\hbar\omega_0+\varepsilon$ is:
\begin{equation}
 \label{eq:Lumin}
 P_\sigma(\varepsilon)
 =\frac{1}{Z}\sum_{N_+}{\frac{p(N_+)\Gamma^2}{(\Delta
 E_\sigma(N,N_+)-\varepsilon)^2+\Gamma^2}}.
\end{equation}

\begin{figure}[!htb]
        \centering
            \subfigure{
                \put(-50,100){\parbox{4.058cm}{\sf \bf a}}{\includegraphics[width=4.058cm]{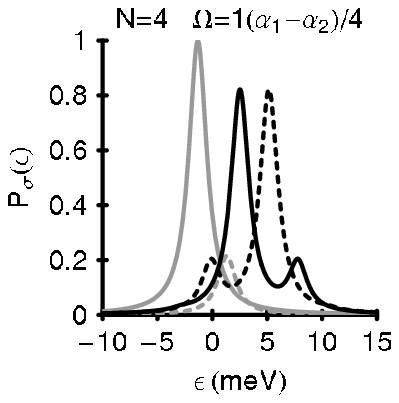}}
                }
            \subfigure{
                \put(-50,100){\parbox{4.058cm}{\sf \bf b}}{\includegraphics[width=4.058cm]{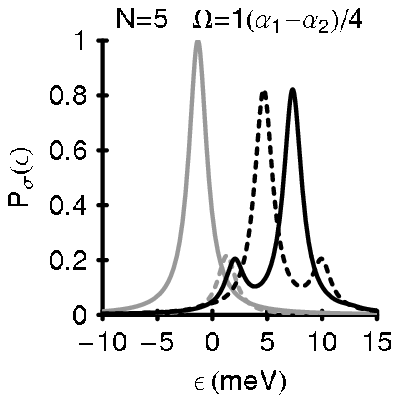}}
                }
            \\
            \subfigure{
                \put(-50,100){\parbox{4.058cm}{\sf \bf c}}{\includegraphics[width=4.058cm]{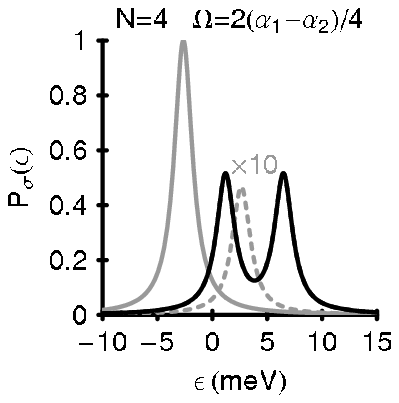}}
                }
            \subfigure{
                \put(-50,100){\parbox{4.058cm}{\sf \bf d}}{\includegraphics[width=4.058cm]{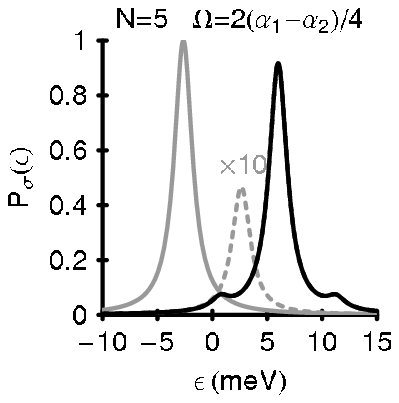}}
                }
            \\
            \subfigure{
                \put(-50,100){\parbox{4.058cm}{\sf \bf e}}{\includegraphics[width=4.058cm]{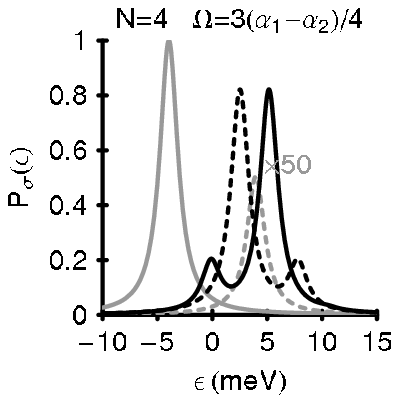}}
                }
            \subfigure{
                \put(-50,100){\parbox{4.058cm}{\sf \bf f}}{\includegraphics[width=4.058cm]{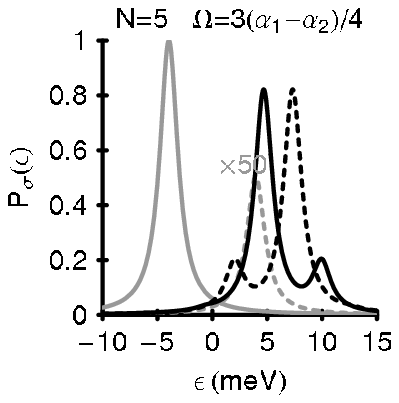}}
                }
            \\
            \subfigure{
                \put(-50,100){\parbox{4.058cm}{\sf \bf g}}{\includegraphics[width=4.058cm]{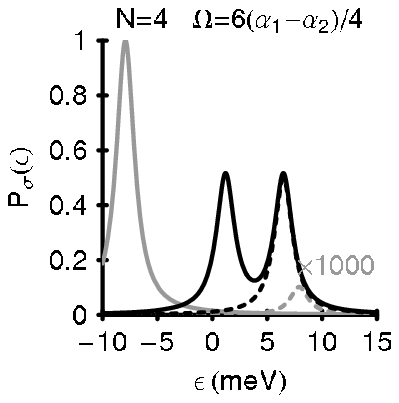}}
                }
            \subfigure{
                \put(-50,100){\parbox{4.058cm}{\sf \bf h}}{\includegraphics[width=4.058cm]{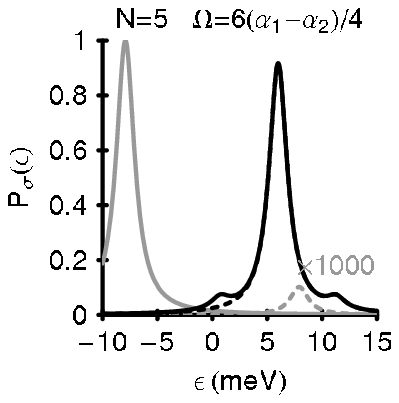}}
                }
    \caption{PL spectra of $\sigma_+$ emission (solid curves) and
    $\sigma_-$ emission (dashed curves) for an even (left) and odd (right) fixed number of
    polaritons for different magnetic fields. To illustrate the screening of the Zeeman splitting,
    grey curves show the spectra in the absence of polariton-polariton interactions.
    We used the parameters: $T=20$K, $\Gamma=1$meV.
    }
    \label{fig:Spectrum}
    \end{figure}

The PL spectra obtained from Eq.~(\ref{eq:Lumin}) are plotted in
Fig.~\ref{fig:Spectrum}.

When $\Omega$ is an integer multiple of $(\alpha_1-\alpha_2)/2$ and
$\Omega<\Omega_c$, peaks in the $\sigma_+$ and $\sigma_-$ emission
overlap, i.e., for special values of the field there is complete
suppression of the Zeeman splitting. From Fig.~\ref{fig:DeltaE} we
note that at these values of $\Omega$ there is a crossing of the
possible $\sigma_+$ and $\sigma_-$ emission energies. Furthermore,
at these values of $\Omega$, there are two initial states that are
highly occupied with equal probabilities. For this reason there are
two peaks in the emission spectra of Figs.~\ref{fig:Spectrum}c (for
both the $\sigma_+$ and $\sigma_-$ cases) and ~\ref{fig:Spectrum}g
(for only the $\sigma_+$ case) appearing with equal intensities.
Above the critical field, $\Omega\geq\Omega_c$, the intensity of the
$\sigma_-$ emission decreases as almost all polaritons are in the
$\sigma_+$ state. The splitting between the $\sigma_+$ and
$\sigma_-$ emission peaks increases above $\Omega_c$ as the emission
energies separate according to Fig.~\ref{fig:DeltaE}.

\section{Photoluminescence for a variable number of polaritons}
 In reality the number of polaritons in the dot can fluctuate due to the
spontaneous and stimulated scattering of polaritons into the dot and
their radiative escape, so that one should average the PL spectra
over $N$ as well. In general, the statistics should depend on the
excitation conditions. Here we assume that the polaritons are at
quasi-equilibrium and one can introduce an effective chemical
potential, $\mu$, so that the distribution is given by
\begin{equation}
 \label{eq:pGibbs}
 p(N,N_+)=Z_G^{-1}e^{\left(\mu N-E(N,N_+)\right)/(k_B T)},
\end{equation}
where $Z_G$ is the grand partition function.

An important effect due to fluctuations of $N$ is the change in the
parity of the polariton number. Polarization steps are expected to
be observed for fluctuating $N$ as well, but they should represent a
superposition of steps for even and odd $N$. This way the width of
each step becomes $(\alpha_1-\alpha_2)/2$, i.e., a half of the width
of the step for the fixed $N$ case. To evidence these ``half-steps"
one should operate with small dots containing a low number of
polaritons. The PL spectra for a fluctuating number of polaritons
are shown in Figs.~\ref{fig:SpectrumN} and
~\ref{fig:SpectrumNDensity}. The left hand column of plots show the
results for the case $T=20$K, which can be compared to the left
column of Fig.~\ref{fig:Spectrum}. For some values of magnetic
fields faint additional peaks can be observed in
Fig.~\ref{fig:SpectrumN} corresponding to the mixing of states with
different numbers of polaritons. The mixing is more obvious for the
case $T=40$K (see the right hand column of plots in
Fig.~\ref{fig:SpectrumN}). Note that, in the case of varying numbers
of particles, $\sigma_+$ and $\sigma_-$ emission spectra still
overlap when $\Omega$ is an integer multiple of
$(\alpha_1-\alpha_2)/2$ and $\Omega<\Omega_c$. Also, at these values
of the field there are pairs of peaks in the emission spectrum
appearing with equal intensities. As in the case of a fixed number
of polaritons, this is caused by two different initial states having
the same probability of occupation. Note that this is in sharp
contrast to what we would expect if there were no interactions, in
which case only single peaks appear in the spectra (since the
emission energy does not depend on the particle number) but with a
large energy splitting and intensity difference between $\sigma_+$
and $\sigma_-$ polarizations.

\afterpage{
    \begin{figure}[H]
        \centering
            \subfigure{
                \put(0,100){\parbox{4.058cm}{\sf \bf a}}{\includegraphics[width=4.058cm]{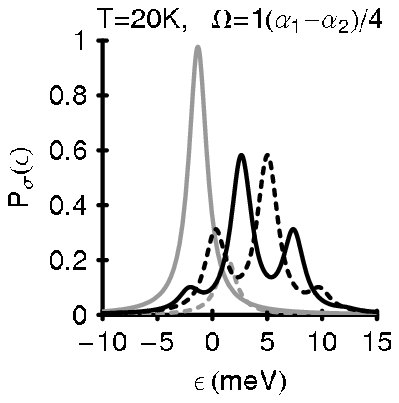}}
                }
            \subfigure{
                \put(0,100){\parbox{4.058cm}{\sf \bf b}}{\includegraphics[width=4.058cm]{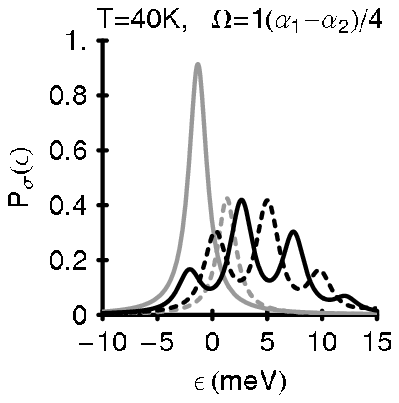}}
                }
            \\
            \subfigure{
                \put(0,100){\parbox{4.058cm}{\sf \bf c}}{\includegraphics[width=4.058cm]{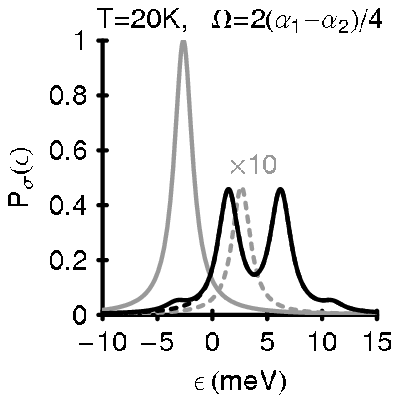}}
                }
            \subfigure{
                \put(0,100){\parbox{4.058cm}{\sf \bf d}}{\includegraphics[width=4.058cm]{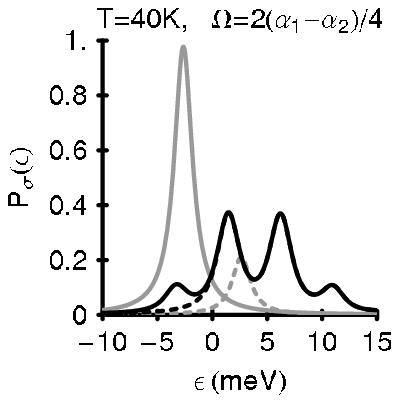}}
                }
            \\
            \subfigure{
                \put(0,100){\parbox{4.058cm}{\sf \bf e}}{\includegraphics[width=4.058cm]{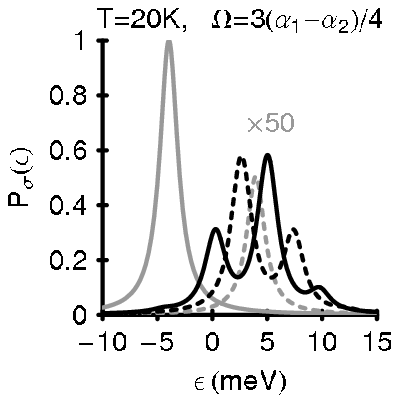}}
                }
            \subfigure{
                \put(0,100){\parbox{4.058cm}{\sf \bf f}}{\includegraphics[width=4.058cm]{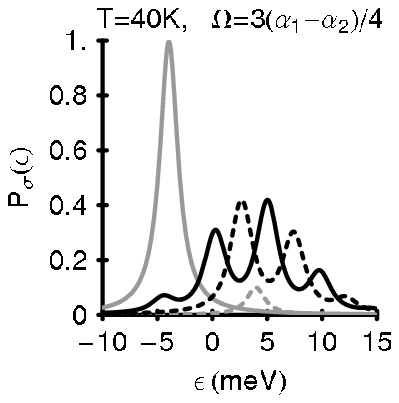}}
                }
            \\
            \subfigure{
                \put(0,100){\parbox{4.058cm}{\sf \bf g}}{\includegraphics[width=4.058cm]{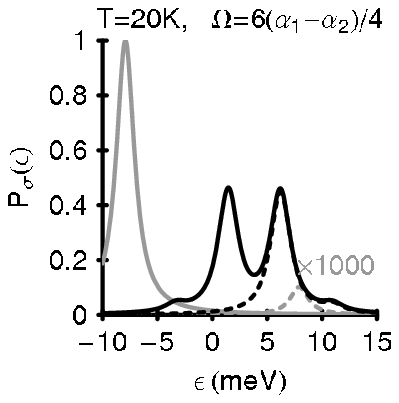}}
                }
            \subfigure{
                \put(0,100){\parbox{4.058cm}{\sf \bf h}}{\includegraphics[width=4.058cm]{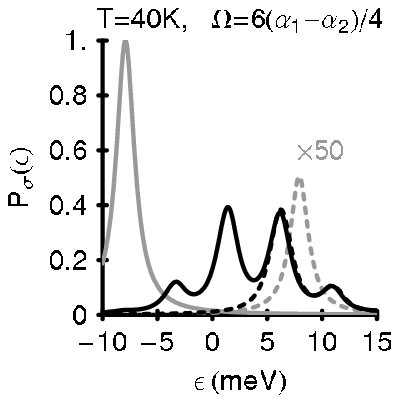}}
                }
                        \\
            \subfigure{
                \put(0,100){\parbox{4.058cm}{\sf \bf i}}{\includegraphics[width=4.058cm]{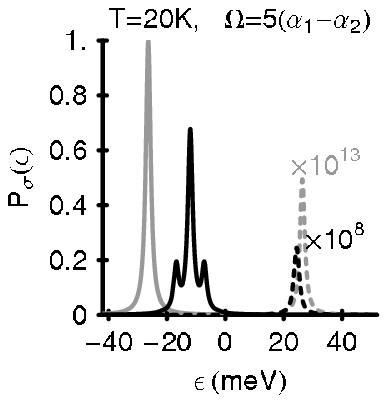}}
                }
            \subfigure{
                \put(0,100){\parbox{4.058cm}{\sf \bf j}}{\includegraphics[width=4.058cm]{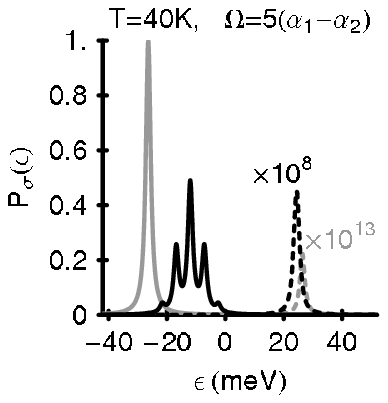}}
                }
    \caption{$\sigma_+$ (solid curves) and $\sigma_-$ (dashed curves) PL spectra for a fluctuating number of polaritons.
    $\mu$ is chosen so that on average there are $4$ polaritons in the dot. Left column: $T=20$K, Right column: $T=40$K. Grey curves show the spectra in the absence
of polariton-polariton interactions. $\Gamma=1$meV.}
    \label{fig:SpectrumN}
    \end{figure}
    \begin{figure}[!htb]
        \centering
            \subfigure{
                \put(-50,100){\parbox{4.058cm}{\sf \bf a}}{\includegraphics[width=4.058cm]{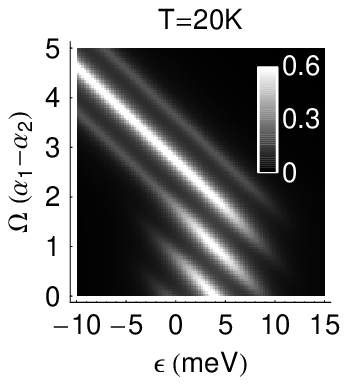}}
                }
            \subfigure{
                \put(-50,100){\parbox{4.058cm}{\sf \bf b}}{\includegraphics[width=4.058cm]{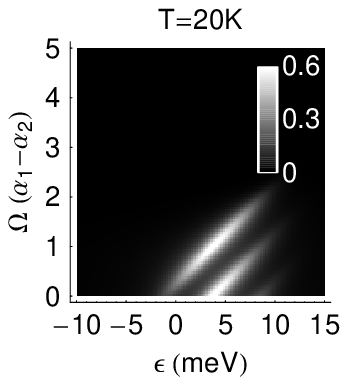}}
                }
            \\
            \subfigure{
                \put(-50,100){\parbox{4.058cm}{\sf \bf c}}{\includegraphics[width=4.058cm]{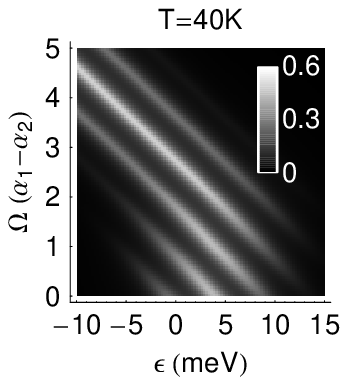}}
                }
            \subfigure{
                \put(-50,100){\parbox{4.058cm}{\sf \bf d}}{\includegraphics[width=4.058cm]{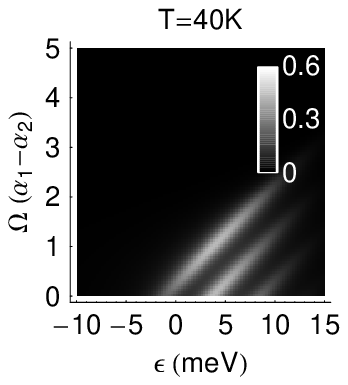}}
                }
    \caption{PL spectra for a a fluctuating number of interacting polaritons. $\mu$ is chosen so that on average there are $4$ polaritons in the dot. Left column: $\sigma_+$ emission, Right column: $\sigma_-$ emission.
    }
    \label{fig:SpectrumNDensity}
    \end{figure}
}

Although there is no well-defined critical field for a fluctuating
number of polaritons, the Zeeman splitting is no longer sufficiently
suppressed by polariton-polariton interactions at high magnetic
fields (see Figs.~\ref{fig:SpectrumN}i,~\ref{fig:SpectrumN}j or
Fig.~\ref{fig:SpectrumNDensity}). Also, at high magnetic fields, the
intensity of the $\sigma_-$ emission decreases as almost all
polaritons are in the $\sigma_+$ state. Since the effect of
polariton-polariton interactions is to suppress the Zeeman
splitting, the intensity of $\sigma_-$ emission is stronger than
that which we would expect if there were no interactions.

It should be noted that there are a relatively small number of peaks
seen in the PL spectra compared to what one could expect to observe
due to the large number of possible transitions that occur in the
case of a fluctuating number of polaritons. This happens because of
the small value of the interaction constant $\alpha_2$ that defines
weak attraction between polaritons with opposite spins. In the
absence of this attraction, the frequency of a luminescence line in
the $\sigma_+$ spectrum is defined solely by the initial number,
$N_+$, of spin-up polaritons and this frequency is independent of
the total polariton number $N$. Similarly, all transitions with a
fixed difference $N-N_+$ contribute to the same line in the
$\sigma_-$ spectrum. The finite value of $\alpha_2$ results in a
fine structure of each peak. This fine structure, however, is not
seen in Fig.~\ref{fig:SpectrumN} due to strong life-time broadening.

\begin{figure}[h]
        \centering
            \subfigure{
                \put(-50,120){\parbox{4.058cm}{\sf \bf a}}{\includegraphics[width=4.058cm]{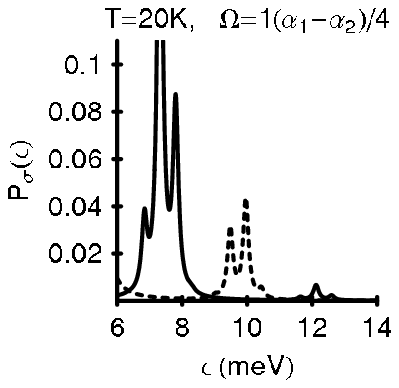}}
                }
            \subfigure{
                \put(-50,120){\parbox{4.058cm}{\sf \bf b}}{\includegraphics[width=4.058cm]{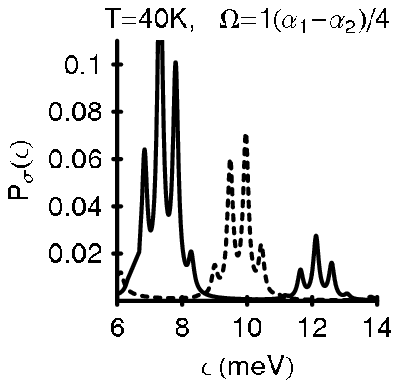}}
                }
    \caption{$\sigma_+$ (solid curves) and $\sigma_-$ (dashed curves) PL spectra for a fluctuating number of polaritons and $\Gamma=0.1$meV.
    $\mu$ is chosen so that on average there are $4$ polaritons in the dot. a) $T=20$K, b) $T=40$K.}
    \label{fig:SpectrumNFine}
    \end{figure}

To show this fine structure, we plotted the PL spectrum for a much
longer polariton life-time in Fig.~\ref{fig:SpectrumNFine}. One can
see the satellite emission around $\epsilon\approx10$meV. The faint
peaks group to form each emission line. In particular, the faint
peaks in the $\sigma_-$ spectra appear due to the transitions from
the states $(N,N_+)=(3,0)$, $(4,1)$, $(5,2)$, $(6,3)$, $(7,4)$, etc.
These states have probabilities of occupation equal to $0.001$,
$0.020$, $0.068$, $0.056$, and $0.012$ respectively for the case
$T=40$K. The emission energies which arise from these states are
separated by $|\alpha_2|$.

\section{Conclusions}
To summarize, we have shown theoretically that, below the critical
magnetic field, governed by the number of polaritons and the
difference between polariton-polariton interaction constants in
triplet and singlet configurations, the circular polarization degree
of a localized polariton condensate increases as a step-like
function of the field. For a fixed number of polaritons the width of
each step is $\Delta B=(\alpha_1-\alpha_2)/(g\mu_B)$. The values of
the field at which steps appear depend on the parity of the total
polariton number, $N$, so that the step width is halved when
fluctuations in $N$ are present. At finite temperatures, the Zeeman
splitting between $\sigma_+$ and $\sigma_-$ emission energies
remains close to zero below the critical magnetic field. The
polarization steps can be evidenced at finite temperatures in
magneto-PL experiments.

TCHL acknowledges support from the EPSRC. YGR acknowledges
DGAPA-UNAM for the support under the grant IN107007. IAS
acknowledges support from the grant of the President of Russian
Federation.



\begin{thebibliography}{10}

\bibitem{Kasprzak2006} J. Kasprzak, M. Richard, S. Kundermann, A. Baas, P. Jeambrun,
 J. M. J. Keeling, F. M. Marchetti, M. H. Szyma\'nska, R. Andr\'e, J. L. Staehli,
 V. Savona, P. B. Littlewood, B. Deveaud, and Le Si Dang,
 Nature \textbf{443}, 409 (2006).

\bibitem{Balili2007} R. Balili, V. Hartwell, D. Snoke, L. Pfeiffer, and K. West,
 Science \textbf{316}, 1007 (2007).

\bibitem{Butov2007} L. V. Butov, Nature {\bf 447}, 540 (2007).

\bibitem{Laussy2006} F. P. Laussy, I. A. Shelykh, G. Malpuech, and A. Kavokin,
 Phys. Rev. B \textbf{73}, 035315 (2006).

\bibitem{Shelykh2006} I. A. Shelykh, Yu. G. Rubo, G. Malpuech, D. D. Solnyshkov, and A. Kavokin,
 Phys. Rev. Lett. \textbf{97}, 066402 (2006).

\bibitem{Rubo2006} Yu. G. Rubo, A. V. Kavokin, and I. A. Shelykh,
Phys. Lett. A \textbf{358}, 227 (2006).

\bibitem{Kasprzak2007} J. Kasprzak, R. Andr\'{e}, Le Si Dang, I. A. Shelykh,
 A. V. Kavokin, Yu. G. Rubo, K. V. Kavokin, and G. Malpuech,
 Phys. Rev. B \textbf{75}, 045326 (2007).

\bibitem{Sanvitto} D. Sanvitto, D. N. Krizhanovskii, D. M. Whittaker, S. Ceccarelli,
 M. S. Skolnick, and J. S. Roberts, Phys. Rev. B \textbf{73}, 241308 (2006).

 \bibitem{Bajoni2007} D. Bajoni, P. Senellart, A Lema\^{i}tre, and J. Bloch,
 Phys. Rev. B, \textbf{76}, 201305 (2007).

\bibitem{Lozovik} See, e.\ g., Yu.E. Lozovik, I. V. Ovchinnikov, S. Yu. Volkov, L. V. Butov,
 and D. S. Chemla, Phys. Rev. B \textbf{65}, 235304 (2002), and references therein.

\bibitem{Savona} See, e.\ g., V. Savona, L. C. Andreani, P. Schwendimann, and A. Quattropani,
 Solid State Commun., \textbf{93}, 733 (1995);
 C. Piermarocchi, A. Quattropani, P. Schwendimann, F. Tassone,
 Phase Transitions \textbf{68}, 169 (1998).

\bibitem{Tignon} J. Tignon, P. Voisin, C. Delalande, M. Voos,
 R. Houdr\'e, U. Oesterle, and R. P. Stanley,
 Phys. Rev. Lett. \textbf{74}, 3967 (1995).

\bibitem{Gibbs} J. D. Berger, O. Lyngnes, H. M. Gibbs, G. Khitrova, T. R. Nelson, E. K. Lindmark,
 A. V. Kavokin, M. A. Kaliteevski, and V. V. Zapasskii,
 Phys. Rev. B \textbf{54}, 1975 (1996).

\bibitem{Ualpha} The constants $\alpha_1$ and $\alpha_2$ are related
to the parameters $U_0$ and $U_1$ used in Ref. \onlinecite{Rubo2006}
as $\alpha_1=U_0/A$ and $\alpha_2=(U_0-2U_1)/A$, where $A$ is the
effective area of polariton dot.

\bibitem{Renucci} P. Renucci, T. Amand, X. Marie, P. Senellart, J. Bloch,
 B. Sermage, and K. V. Kavokin,
 Phys. Rev. B \textbf{72}, 075317 (2005).

\bibitem{Shapira} Y. Shapira, in
 \emph{Semimagnetic Semiconductors and Diluted Magnetic Semiconductors},
 edited by M. Averous and M. Balkanski (Plenum, New York, 1991), p.121.

\bibitem{Rubo1997} Yu. G. Rubo, M. F. Thorpe, and N. Mousseau, Phys.
Rev. B \textbf{56}, 13094 (1997).

\end{thebibliography}
\end{document}